\documentstyle[12pt,prb,aps,psfig]{revtex}
 \begin{document}
 \draft
\title{Symmetry of Quantum Phase Space in a 
Degenerate Hamiltonian System}
 \author{G. P. Berman\dag, V. Ya. Demikhovskii\ddag,  
  \\and D. I. Kamenev\dag\hspace{0.2cm} (E-mail: kamenev@cnls.lanl.gov) 
 \footnote{on leave from Nizhny Novgorod State University, Nizhny Novogorod,
 603600, Russia}}
 \address{\dag Theoretical Division and CNLS, Los Alamos National Laboratory, Los Alamos, NM 87545}
 \address{\ddag Nizhny Novgorod State University, Nizhny Novogorod,
 603600, Russia}
 \maketitle

\vspace{1cm}
\begin{abstract}
Using Husimi function approach, we study the ``quantum phase space'' of a harmonic oscillator 
interacting with a plane monochromatic wave. We show that in the regime of weak chaos, 
the quantum system has the same symmetry as the classical system. 
Analytical results agree with the results of numerical calculations.
\end{abstract}
\vspace{1cm}
It is known that the phase space of a
classical harmonic oscillator weakly interacting with a plane monochromatic wave    
possesses an interesting symmetry. (See, for example, \cite{Z} and references therein.) In the 
case of exact resonance, $\mu\omega=\Omega$ ($\mu=1,2,\dots$), 
between the wave (with the frequency $\Omega$)  
and the harmonic oscillator (with the oscillation frequency $\omega$), 
and under the condition $\epsilon\ll 1$ 
(where $\epsilon$ is a dimensionless perturbation parameter), 
the classical 
phase space consists of an infinite number of 
resonant cells with the symmetry $2\mu$. An example of a corresponding phase space with $\mu=4$ is shown 
in Fig. 1. At the center of
each cell there is an elliptic stable point. The
particles move in the phase space around this point along the
closed trajectories. The cells are separated from each other  
by the separatrices which are 
schematically shown in Fig. 1 by dashed lines. 
These separatrices form in the phase space  
an unlimited net. The net is covered by the stochastic layers 
forming the infinite stochastic web. When the  perturbation parameter, $\epsilon$, 
is small the web width is exponentially thin \cite{Z}. However, if the particle 
is initially placed inside a stochastic region, it can travel throughout the
web and gain energy, even for an arbitrarily small perturbation parameter, $\epsilon$.
The existence of the crystalline and quasi-crystalline symmetries of the classical phase space, and stochastic web differ significantly this system from classical nonlinear systems with chaotic behavior.\cite{Z}
These interesting properties of the classical harmonic oscillator in  
a monochromatic wave  motivated our studies of the corresponding properties in the quantum system.

The quantum harmonic oscillator interacting with  a monochromatic wave is 
described by the Hamiltonian, 
\begin{equation}
\label{ham}
\hat H=\frac{\hat p^2}{2m}+\frac{m\omega^2x^2}2+
\frac\varepsilon k\cos(kx-\Omega t)=\hat H_0+\hat V(x,t),
\end{equation}
where $\hat H_0$ is the Hamiltonian of the harmonic oscillator; $\hat V(x,t)$, is the 
interaction Hamiltonian;  
$\varepsilon/k$ and $k$ are, respectively, the amplitude and
the wave vector of the wave,
$x$ and $\hat p$ are the coordinate and the momentum operators of the particle, 
$m$ is the mass of the particle. 
The Hamiltonian (1) appears, for example, when analyzing the stability of an ion in a 
linear ion trap in the field of two  laser beams with close frequencies \cite{ber1}.
The dynamics of the quantum system described by the 
Hamiltonian (\ref{ham}) is controlled by four parameters:
the resonance number, $\mu$; the detuning from the exact resonance: 
$\delta\omega=\mu\omega-\Omega$; the dimensionless perturbation 
parameter: $\epsilon=\varepsilon k/m\omega^2$; and
the dimensionless Planck constant: $\hbar_0=(\hbar k^2)/(m\omega)$. (For the system 
considered in \cite{ber1}, the Lamb-Dicke parameter, $\eta$, is related to 
$\hbar_0$: $\hbar_0=2\eta^2$).  
Influence of these parameters on the dynamics of 
quantum system was in detail considered in Refs. \cite{ber1,1,2,3}  

In this paper, we 
study the ``quantum phase space'' of a system described by the 
Hamiltonian (\ref{ham}) for the case of exact resonance,
$\delta\omega=0$ (when the number of the resonance  
cells is infinite) and under the condition: $\epsilon\ll 1$ (when the chaotic layers, 
covering the separatrix net, are exponentially thin). 
We investigate the structure of the ``quantum phase space''. We show that quantum system 
possesses the same symmetry as the classical system.    
The structure of the quantum system with the Hamiltonian (\ref{ham})
is characterized by the Floquet 
\begin{figure}[tb]
\begin{center}
\psfig{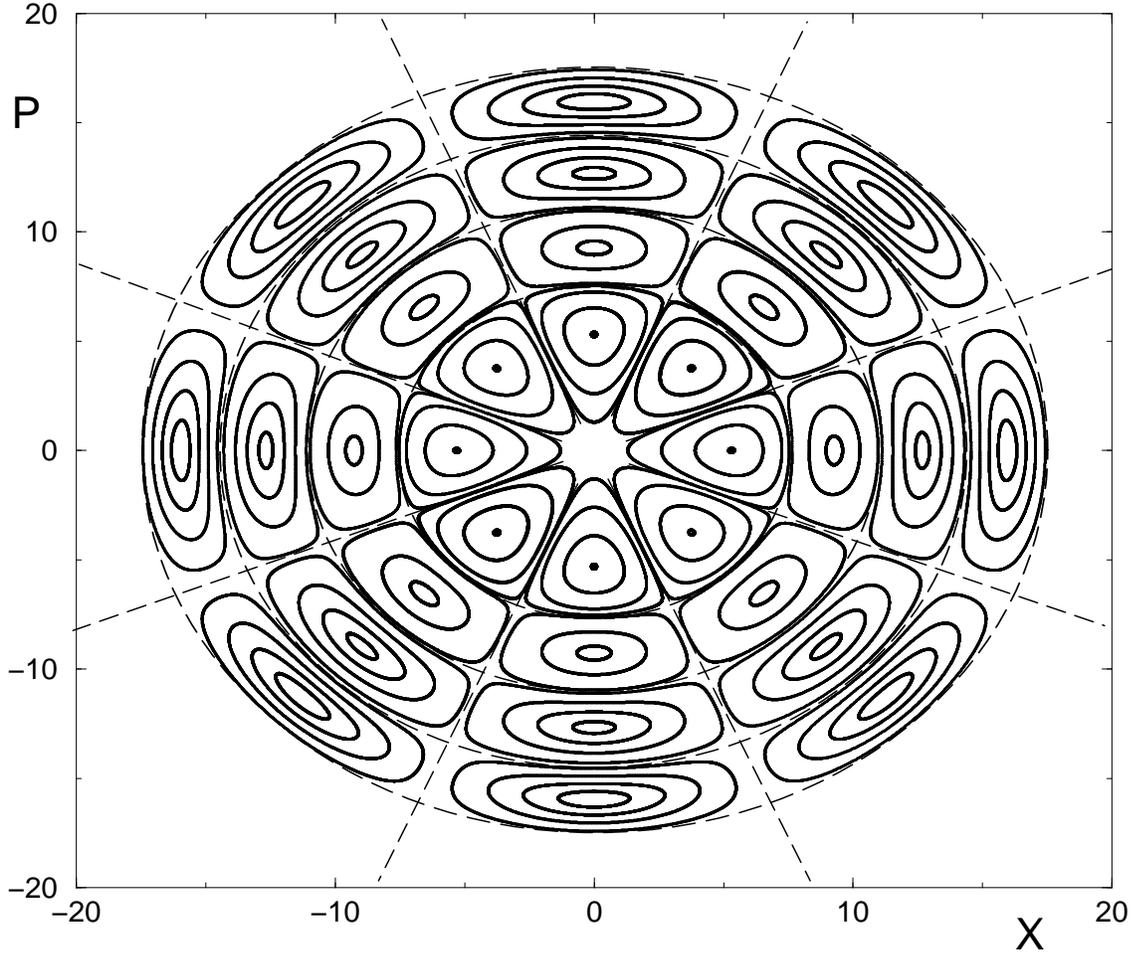}
\vspace{0.3cm}
\caption{The classical phase space for a harmonic oscillator in a monochromatic 
wave, for: $\delta\omega=0$ (exact resonance),
$\mu=4$, $\epsilon=0.05$.}
\end{center}
\label{2.2}
\end{figure}

\noindent
states (or quasienergy states)
found in Refs. \cite{1,2,3}.
In order to build the phase space for the
quantum system we use 
the Husimi functions of the Floquet states.
The Husimi function for the wave 
function $\psi(x,t)$    
is defined as the projection  
of $\psi(x,t)$ on the coherent wave packet, $\chi(x;\,X,\,P),$
with the maximum at the point ($X,\,P$) \cite{husimi},  
\begin{equation}
\label{hus}
\Phi(X,\,P,\,t)=\frac 1{2\pi}|<\psi(x,\,t)|\chi(x;\,X,\,P)>|^2.
\end{equation}
The Husimi function, $\Phi(X,P,t)$,
defines the probability of finding a quantum particle characterized
by the wave function $\psi(x,t)$ at the point $(X\,,P)$ of the
``quantum phase space''. The cross-sections of the Husimi function are
the lines of equal probability of finding the quantum particle.
Below, these lines for Husimi functions are compared with
the trajectories in classical phase space.

Namely, we analyze the structure of the Husimi functions 
of the quasienergy states and compare them with the structure of the 
classical phase space. First, we present some general formulas
which will be used to investigate the system
described by the Hamiltonian (\ref{ham}).    
It is convenient to decompose the coherent state, $\chi(x;X,\,P)$,  
into the complete set of harmonic oscillator eigenstates,
\begin{equation}
\label{coh}
\chi(x;\,X,\,P)=\exp\left(-\frac{X^2+P^2}{4\hbar_0}\right)
\sum_{m=0}^\infty \frac{(X+iP)^m}{\sqrt{(2\hbar_0)^mm!}}\psi_m(x),
\end{equation}
where $\psi_n(x)$ is the $n$-th eigenfunction of the harmonic oscillator 
with the Hamiltonian $\hat H_0$. In Eq. (\ref{coh}) we used the dimensionless  
coordinate ($X=kx$) and the dimensionless momentum
($P=pk/m\omega$).  
We use the same basis to represent 
of the wave function, $\psi(x,t)$,    
\begin{equation}
\label{sho_decomposition}
\psi(x,t)=\sum_{n=0}^\infty C_n(t)\psi_n(x)
\exp\left[-i\omega t\left(n+\frac 12\right)\right].
\end{equation}
The structure of the Husimi function (\ref{hus}) 
is completely defined by the coefficients $C_n(t)$,
\begin{equation}
\label{coh1}
\Phi(X,\,P,\,t)={\exp\left(-\frac{X^2+P^2}{2\hbar_0}\right)\over 2\pi}
\left|\sum_{m=0}^\infty C^*_m(t)\frac{(X+iP)^m}{\sqrt{(2\hbar_0)^mm!}}
\exp(im\omega t)\right|^2.
\end{equation}
It is convenient to use cylindric coordinates,
\begin{equation}
\label{r_phi}
X=r\cos\varphi,\qquad P=r\sin\varphi,
\end{equation}
where $r=\sqrt{X^2+P^2}$ and $\varphi={\rm arctg}(P/X)$. In these
variables, the Husimi function (\ref{coh1}) is,
\begin{equation}
\label{hus_dyn}
\Phi(r,\,\varphi,\,t)={\exp\left(-\frac{r^2}{2\hbar_0}\right)\over 2\pi}
\left|\sum_{m=0}^\infty C^*_m(t)\frac{r^me^{im\varphi}}{\sqrt{(2\hbar_0)^mm!}}
\exp(im\omega t)\right|^2.
\end{equation}

Since the perturbation, $V(x,t)$, in (1) is periodic in time, one can use 
Floquet theory and write the solution of the non-stationary
Schr\"odinger equation as, 
\begin{equation}
\label{qe_ef}
\psi _q(x,t)=\exp(-iE_qt/\hbar)U_q(x,t),
\end{equation}
where $U_q(x,t)=U_q(x,t+T)$ is a time-periodic function whose period 
is $T=2\pi/\Omega$. 
The index $q$ labels the quasienergy (QE) states. It is convenient 
to use the complete set of harmonic oscillator eigenfunctions to 
represent  the function $U_q(x,t)$, 
\begin{equation}
\label{QE_decomposition}
U_q(x,t)=\sum_{n=0}^\infty C_n^q(t)\psi_n(x),
\end{equation}
where the expansion coefficients, 
$C_n^q(t)=C_n^q(t+T)$, are time-periodic functions. 
Using Eqs. (\ref{hus_dyn})-(\ref{QE_decomposition})
we can rewrite the Husimi function of the QE state as,
\begin{equation}
\label{hus_qe}
\Phi_q(r,\,\varphi,\,sT)={\exp(-\frac{r^2}{2\hbar_0})\over 2\pi}
\left|\sum_{m=0}^\infty C^{q*}_m\frac{r^me^{im\varphi}}{\sqrt{(2\hbar_0)^mm!}}
\exp\left(2\pi ims\frac{\omega}{\Omega}\right)\right|^2,
\end{equation}
\noindent
where $s=0,\,1,\,2,\dots$.
The QE states of the monochromatically perturbed harmonic oscillator 
were studied in detail in a series of papers \cite{1,2,3}, using 
degenerate resonance perturbation theory for the Floquet
states. In particular, the quantum regimes corresponding to
regular motion and to the case of weak chaos in the classical 
phase space were investigated. As it was shown in Ref. \cite{1}, the Hilbert space 
of the quantum system breaks up to  some approximation into the  dynamically independent 
regions --- quantum resonance cells, each of them with its own set of
QE states. In the zeroth order (resonance) approximation, the QE functions 
and the QE spectrum of each cell are almost independent.   
Near the top and bottom of the QE spectrum of an individual cell, 
the QE states are the states of an effective harmonic oscillator. The QE levels are 
equally spaced, with a separation $\hbar\tilde\omega$ between the
levels. The frequency, 
$\tilde\omega$, in the quasiclassical limit coincides with 
the frequency of small oscillations near the center of the 
resonance in phase space. In this paper, we consider only  
two extreme QE states of an individual cell -- extreme upper and 
extreme lower QE states, called the ``QE ground states''. 
Thus, each quantum resonance cell has two QE ground states.
The QE functions, $C_n^q$, and the QE levels, $E_q$, of the    
ground states of each individual cell are connected by the relations \cite{2},  
\begin{equation}
\label{transform}
E_q \rightarrow -E_q, \qquad C_{\mu m}^q \rightarrow (-1)^mC_{\mu m}^q,
\end{equation}
associated with the transformation: $x \rightarrow -x$
in Eqs. (\ref{sho_decomposition}),(\ref{QE_decomposition}).       

Below we analyze the structure of the Husimi 
functions of the QE ground states. The upper QE function is a
Gaussian wave packet, 
\begin{equation}
\label{coh_n}
C_n^{q_e}=\Gamma\exp\left(-{(n-n_e)^2\over 2a_e^2} \right),
\end{equation}
where $\Gamma$ is the normalization factor, and $n_e$ is the position
of the maximum of the QE wave packet in the Hilbert space
which corresponds to the quantized radius of the elliptic stable point:
$r_e=\sqrt{2n_e\hbar_0}$ (see Ref. \cite{2}). The width
of the wave packet, $a_e$, in Eq. (\ref{coh_n}) was defined
in Ref. \cite{2} in the form, 
\begin{equation}
\label{width_n}
a_{e}=\left({g_\mu(n_e)\over g_\mu''(n_e)}\right)^{1/4},
\end{equation}
were the function $g_\mu(n)$ is expressed in terms of the matrix element:
$g_\mu(n)=<\psi_n|\cos(kx)|\psi_{n+\mu}>$. In the quasiclassical
region of parameters, Eq. (\ref{width_n}) can be expressed through
the half width in action of the classical resonance cell, $\Delta I$ 
(expression for the value of $\Delta I$ see for example 
in Ref. \cite{L_b}),
\begin{equation}
\label{width_n1}
a_{e}=\left({(r_e)^2\over \hbar_0^2}\left|
{J_\mu(r_e)\over J_\mu''(r_e)}\right|\right)^{1/4}=
\left(\frac{\Delta I_e}{\sqrt{2}\hbar_0}\right)^{1/2},
\end{equation}
where the prime indicates differentiation with respect to the argument.
For example, for $\mu=1$ we have: $a_e=r_e/\{\hbar_0^2[(r_e)^2-1]\}^{1/4}$.
The boundaries of the quantum cells are given by the zeroes of the 
function $g_\mu(n)$. 
As was shown in Ref. \cite{1}, the function $g_\mu(n)$ is 
proportional to the Bessel function, $J_\mu$, of order $\mu$: 
$g_\mu(n)\sim J_\mu(\sqrt{2n\hbar_0})$. So,  the number of  
levels in the individual cell is proportional to $\hbar_0$. 
Thus, the ratio:
(the packet's width in $n$)/(the cell's width in $n$) is proportional to
$\sqrt \hbar_0$, and in the quasiclassical limit the relative width of the
QE ground state tends to zero. 

The Husimi representation allows one to construct the QE eigenstates
in the quantum phase space. The simplest case is the Husimi function
of a single harmonic oscillator state: $C_n=\delta_{n,n_0}$, which due to
Eq. (\ref{hus_dyn}) has the form,
\begin{equation}
\label{hus_lan}
\Phi^{(n_0)}(r,\,\varphi,\,t)={\exp\left(-\frac{r^2}{2\hbar_0}\right)\over 2\pi}
{r^{2n_0}\over (2\hbar_0)^{n_0}n_0!}.
\end{equation}
This expression has its maximum at
$r_0=\sqrt{2n_0\hbar_0}$. The definite value of $n_0$ corresponds
to the definite value of the action: $I_0=\hbar_0 n_0$. 
Due to the fundamental uncertainty
relation, the phase, $\varphi$, of this state is indefinite. The Husimi
function is independent of the phase, $\varphi$, and
looks like a round hump.

In agreement with Eqs. (10) and (12), the Husimi function of the ground QE state is,
\begin{equation}
\label{phi_q}
\Phi_{q_e}(r,\,\varphi,sT)\equiv\Phi_{e}(r,\,\varphi,s)=
{\exp\left(-\frac{r^2}{2\hbar_0}\right)\over 2\pi}|\Gamma|^2
\left|\sum_{m=0}^\infty\frac{r^me^{im\left(\varphi+\frac{2\pi s}\mu\right)}}{\sqrt{(2\hbar_0)^mm!}}
\exp\left(-{(m-n_e)^2\over 2a_e^2}\right)\right|^2.
\end{equation}
Only $\Delta m\sim a_e$ terms with $|m-n_e|\le 2a_e$ effectively
contribute to the sum on the right-hand side of Eq. (\ref{phi_q}), and
one can neglect all other terms. Then, Eq. (\ref{phi_q}) becomes,
\begin{equation}
\label{phi_q1}
\Phi_e(r,\,\varphi,s)={\exp\left(-\frac{r^2}{2\hbar_0}\right)\over 2\pi}
{r^{2n_e}\over (2\hbar_0)^{n_e}n_e!}|\Gamma|^2
\left|\sum_{n=-\Delta m}^{\Delta m}
\frac{r^ne^{in\left(\varphi+\frac{2\pi s}\mu\right)}}{\sqrt{(2\hbar_0 n_e)^n}}
\exp\left(-{n^2\over 2a_e^2}\right)\right|^2,
\end{equation}
where we assumed: $n_e\gg 1$, so that, 
\begin{equation}
\label{appr}
(n_e+m)!\simeq n_e!n_e^m.
\end{equation}
The double sum in Eq. (\ref{phi_q1}) can be rewritten as,
$$
\sum_{n,m=-\Delta m}^{\Delta m}
\frac{r^{n+m}e^{i(n-m)\left(\varphi+\frac{2\pi s}\mu\right)}}{\sqrt{(2\hbar_0 n_e)^{n+m}}}
\exp\left(-{n^2+m^2\over 2a_e^2}\right)=
$$
$$
\sum_{j=-2\Delta m}^{2\Delta m}\left(\frac r{\sqrt{2\hbar_0 n_e}}\right)^j
\exp\left(-{j^2\over 4a_e^2}\right)
\sum_{k=-2\Delta m}^{2\Delta m}e^{ik\left(\varphi+\frac{2\pi s}\mu\right)}
\exp\left(-{k^2\over 4a_e^2}\right),
$$
where $j=n+m$, $k=n-m$.
Thus, by using the approximation (\ref{appr}) we find
that the Husimi function of the
extreme QE state,  can be factored,
\begin{equation}
\label{phi_q2}
\Phi_e(r,\,\varphi,s)=\gamma(r)\xi(\varphi,s).
\end{equation}
In Eq. (19),
\begin{equation}
\label{hus_r}
\gamma(r)={e^{-\frac{r^2}{2\hbar_0}}\over 2\pi}
{r^{2n_e}|\Gamma|^2\over (2\hbar_0)^{n_e}n_e!}
\sum_{j=-2\Delta m}^{2\Delta m}\left(\frac r{\sqrt{2\hbar_0 n_e}}\right)^j
\exp\left(-{j^2\over 4a_e^2}\right),
\end{equation}
\begin{equation}
\label{hus_phi}
\xi(\varphi,s)=\sum_{k=-2\Delta m}^{2\Delta m}e^{ik\left(\varphi+\frac{2\pi s}\mu\right)}
\exp\left(-{k^2\over 4a_e^2}\right).
\end{equation}

We now find the coordinates of maxima of $\Phi_e(r,\,\varphi)$.
Suppose that each maximum of the Husimi function corresponds to the
stable elliptic point at the center of a resonance cell.
Maximum of $\Phi_q(r,\,\varphi)$ in $r$ is defined from the equation,
$$
\frac d{dr}\gamma(r)=\left[\frac d{dr}{e^{-\frac{r^2}{2\hbar_0}}\over 2\pi}
{r^{2n_e}|\Gamma|^2\over {2\hbar_0}^{n_e}n_e!}\right]
\sum_{j=-2\Delta m}^{2\Delta m}\left(\frac r{r_e}\right)^j
\exp\left(-{j^2\over 4a_e^2}\right)+
$$
\begin{equation}
\label{dhus_r}
{e^{-\frac{r^2}{2\hbar_0}}\over 2\pi}{r^{2n_e}|\Gamma|^2\over {2\hbar_0}^{n_e}n_e!}
\sum_{j=-2\Delta m}^{2\Delta m}
\frac j{r_e}\left(\frac r{r_e}\right)^{j-1}
\exp\left(-{j^2\over 4a_e^2}\right)=0,
\end{equation}
When $r=r_e$, both sums in Eq. (\ref{dhus_r}) are zero:
in the first term, the derivative is equal
to zero as follows from Eq. (\ref{hus_lan}); in the second term, the sum is equal
to zero, and the value $r_e$ can be considered as the radius of
the center of the quantum resonance cell in the quantum phase space.

We now find the maxima of $\xi(\varphi,s)$. It is convenient
to present this function in the form,
\begin{equation}
\label{hus_phi_ell}
\xi(\varphi)=1+2\sum_{m=1}^{2\Delta m/\mu}\cos(\mu m\varphi)
\exp\left(-{(\mu m)^2\over 4a_e^2}\right),
\end{equation}
where we took into account that in the resonance 
approximation the particle can populate only states with 
the numbers: $k=\mu m$ (see Ref. \cite{1}). 
All terms in the sum on the right-hand side of Eq.
(\ref{hus_phi_ell}) decrease in absolute values as $m$ increases. 
 Then, the extrema of the function $\xi(\varphi)$ is defined by the
the extrema of the term with $m=1$.
When $\mu=1$ there is one maximum at $\varphi=0$; when
$\mu=2$ there are two maxima at $\varphi=0$ and $\varphi=\pi$. In general
case the function $\xi(\varphi)$ has $\mu$ maxima.

The extreme lower QE function is related to the extreme upper one
by the transformation (\ref{transform}), which is convenient to rewrite
in the form: $C_{\mu m}^q\rightarrow\exp(-i\pi m)C_{\mu m}^q$.
The function $\xi_{lower}(\varphi)$ of the lower ground QE state is,
\begin{equation}
\label{hus_phi_rev}
\xi_{lower}(\varphi)=1+2\sum_{m=1}^{2\Delta m/\mu}\cos[(\mu\varphi-\pi)m]
\exp\left(-{(\mu m)^2\over 4a_e^2}\right).
\end{equation}
The maxima of the function $\xi(\varphi)$ in Eq. (\ref{hus_phi_ell}) correspond to
minima of $\xi_{lower}(\varphi)$ in Eq. (\ref{hus_phi_rev}), and vice versa. Thus,
for $\mu=1$ the function $\xi_{lower}(\varphi)$ has a maximum at
$\varphi=\pi$; at $\mu=2$ there are two maxima at
$\varphi=\pm\pi/2$ and so on. In general, 
the Husimi functions of the 
two QE ground states have $2\mu$ maxima with the radius $r=r_e$. Each
maximum is situated at

\begin{figure}[tb]
\begin{center}
\psfig{file=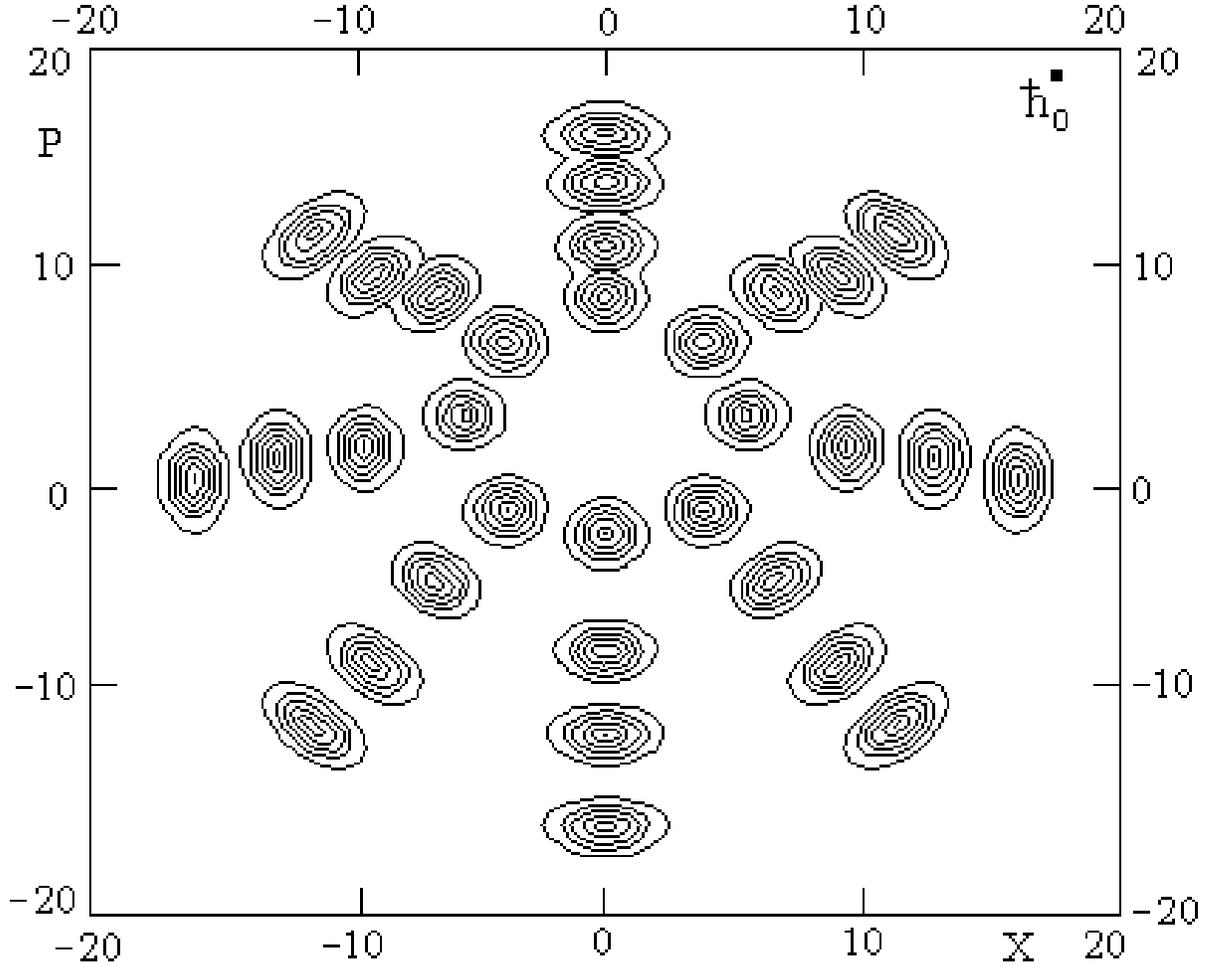,width=18.0cm}
\vspace{0.7cm}
\caption{Contour plots of the Husimi functions for 
the exact resonance case, with the resonance number $\mu=4$; 
$\hbar_0=0.12$; $\epsilon=0.002$.}
\end{center}
\label{3.7}
\end{figure}

\noindent
the center of a quantum resonance cell, so
that the quantum phase space 
has the same symmetry as the classical phase
space. For $\mu=4$, the symmetry of the Husimi function, 
shown in Fig. 2,
is the same as the symmetry of the classical phase space in Fig. 1.
A similar result was demonstrated numerically in Ref. \cite{2} for $\mu=1$. As one can see from Fig. 2, in agreement with Eq. (11), the quantum phase space is
symmetric with respect to the substitution: $X\rightarrow -X$. 
However, there is no exact symmetry with 
respect to the transformation: $P\rightarrow -P$. The reason is 
presumably related to our approximation (\ref{appr})  
which leads to separation of variables in 
Eq. (\ref{phi_q2}). This approximation  is more valid for the ``quasiclassical cells'' 
with 
$n\gg 1$ than for ``quantum  cells'',
for which the value of $n$ is not large.  
One can see from Fig. 2,  that ``quasiclassical cells'' are more symmetrical than 
the ``quantum cells'', and the structure of the ``quasiclassical cells''  is close 
to the structure of the classical cells shown in Fig. 1.
This symmetry of the quantum phase space differs this system from quantum chaotic systems with critical threshold to global chaos.\cite{rp} 

In summary, the correspondence between the symmetry of the Husimi functions 
of the QE ground states and the symmetry of the classical phase space
has been demonstrated for a degenerate system both analytically and numerically.  

\section{Acknowledgments}
We are thankful to D.F.V. James and G.D. Doolen for useful discussions. 
The work of V.Ya.D. and D.I.K. 
was partly supported by the Russian Foundation for Basic
Research (Grants No. 98-02-16412 and No. 98-02-16237).
Work at Los Alamos National Laboratory was partly supported by the National 
Security Agency, and by the Department of Energy under contract W-7405-ENG-36.


\begin{references}
\bibitem{Z} 
G.M. Zaslavsky, R.Z. Sagdeev, D.A. Usikov,
and A.A. Chernikov,
{\it Weak Chaos and Quasi-Regular Patterns}, Cambridge Univ. Press,
Cambridge, 1991.
%
\bibitem{ber1}
G.P. Berman, D.F.V. James, R.J. Hughes, M.S. Gulley, M.H. Holzscheiter, G.V. L\'opez, quant-ph/9903063.
%
\bibitem{1} 
V.Ya. Demikhovskii, D.I. Kamenev, and G.A.
    Luna-Acosta, {\it Phys. Rev. E}, {\bf 52} (1995), 3351.
%
\bibitem{2} 
V.Ya. Demikhovskii, D.I. Kamenev,
    {\it Phys. Lett. A}, {\bf 228} (1997), 391.
%
\bibitem{3} 
V.Ya. Demikhovskii, D.I. Kamenev, and G.A.
    Luna-Acosta, {\it Phys. Rev., E}, {\bf 59} (1999), 294.
%
\bibitem{husimi}
K. Husimi, {\it Proc. Phys. Math. Soc. Jpn.}, {\bf 22} (1940) 264.
%
\bibitem{L_b} 
A.J. Lichtenberg and M.A. Lieberman,
     {\it Regular and Stochastic Motion}, Springer, New York, 1983, Ch. 2.
\bibitem{rp}
G. Radons and R.E. Prange, {\it Phys. Rev. Lett.}, {\bf 61}, (1988) 1691.
\end{references}
\end{document}